\documentclass[reprint,superscriptaddress,amsmath,amssymb,aps,notitlepage,longbibliography,floatfix,nofootinbib,twocolumn]{revtex4-2}

\usepackage{graphicx}
\usepackage{dcolumn}
\usepackage{bm}
\usepackage{natbib}

\usepackage{color}
\usepackage{hyperref}
\hypersetup{colorlinks=true,linkcolor=blue,filecolor=blue,urlcolor=blue,citecolor=blue}
\usepackage{hyperref}
\usepackage[mathlines]{lineno}

\usepackage{multirow}
\usepackage{booktabs} 
\usepackage{threeparttable}

\begin{document}%
\begin{sloppypar}

\title{
GW241011 and GW241110: \\
Hints of Hierarchical Mergers from the Merger Entropy Index 
}

\author{Guo-Peng Li}
\affiliation{
Department of Astronomy, School of Physics and Technology, Wuhan University, Wuhan 430072, China}

\author{Xi-Long Fan}
\email[Corresponding author: Xi-Long Fan, ]{xilong.fan@whu.edu.cn}
\affiliation{
Department of Astronomy, School of Physics and Technology, Wuhan University, Wuhan 430072, China}

\date{\today}

\newcommand{\Msun}{M_{\odot}}

\begin{abstract}
GW241011 and GW241110 both exhibit extremely asymmetric masses, high primary spins, and significant spin–orbit misalignment, which challenge the formation of first-generation binary black hole mergers formed from stellar collapse. This implies that these two gravitational wave events might originate from the hierarchical merger mechanism, with at least one of the black holes being the remnant of a previous merger. Here we investigate the origin of hierarchical mergers for GW241011 and GW241110 using the merger entropy index which measures the efficiency of entropy transfer for binary black hole mergers in general relativity. We find that GW241011 is consistent with hierarchical mergers in dense star clusters. The origin of GW241110 remains under debate due to its large distribution uncertainty, which leads to method-dependent inference and should be taken into account when interpreting this event in terms of hierarchical mergers.
\end{abstract}
\maketitle

\section{Introduction}

Recently, the LIGO-Virgo-KAGRA (LVK) Collaboration reported a pair of gravitational-wave (GW) events discovered in late 2024, GW241011\_233834 (hereafter GW241011) and GW241110\_124123 (hereafter GW241110), identified as asymmetric and high-spin binary black hole (BBH) mergers~\citep{2025ApJ...993L..21A}.
In particular, GW241011 with a high signal-to-noise ratio (SNR) of $\sim$$36$ exhibits precisely measured source properties: the mass ratio and primary dimensionless spin magnitude of GW241011 are well constrained to $q=0.30^{+0.09}_{-0.08}$ and $\chi_1=0.7^{+0.09}_{-0.09}$, respectively, with the primary mass of $m_1=19.6^{+3.6}_{-2.5}\,\Msun$ (see Table\,1 in \citet{2025ApJ...993L..21A}).
GW241110 has similar characteristics to GW241011, however, it exhibits larger uncertainties due to its lower SNR: $m_1=17.2^{+5.0}_{-4.4}\,\Msun$, $q=0.45^{+0.32}_{-0.17}$, and $\chi_1=0.61^{+0.33}_{-0.40}$, for which equal mass ratios and low primary spins cannot be fully excluded.
Both events have non-negligible primary spin–orbit misalignment, with the misalignment angle of GW241011 being $\sim$$30^\circ$ and that of GW241110 being more than $110^\circ$.

The extreme asymmetric masses, high primary spins, and significant spin–orbit misalignment of GW241011 and GW241110 challenge the formation of first-generation (1G) BBH mergers formed from the collapse of stars, 
either originated from isolated binary evolution~\citep{2021hgwa.bookE..16M,2022PhR...955....1M,2016Natur.534..512B,2017NatCo...814906S,2019ApJ...881L...1F,2021ApJ...920...81S,2021ApJ...920L..13B,2022ApJ...928..163H,2024ApJ...965..177W,2024ApJ...974..211Z,2024arXiv241203461B,2019ApJ...887L..36M,2020MNRAS.498.4924M,2022ApJ...936..184M,2022MNRAS.514.4246R,2024ApJ...965..148X,2024ApJ...972L..19S,2025arXiv251019249S}
or dynamical formation~\citep{2016MNRAS.460.3494S,2018ApJ...856..140H,2019ApJ...877...87Z,2020ApJ...894...15B,2020ApJ...901..125D,2021ApJ...917...76W,2023Univ....9..138A,2024ApJ...971L..38K,2024MNRAS.534.1634L,2019PhRvD.100d3027R,2018MNRAS.481.4775D,2020MNRAS.492.2936A,2021ApJ...921L..43Z,2024MNRAS.531..739G,2025MNRAS.538..639B,2023A&A...673A...8D,2021ApJ...907L..20T,2022Natur.603..237S,2022PhRvD.105f3006L,2022A&A...666A.194L,2022arXiv220406002M,2023ApJ...944L..42L,2023MNRAS.524.6015L,2023arXiv230312539G,2025ApJ...979L..27L,2025PhRvD.112f3019D,2025ApJ...983..114W,2025ApJ...993..139W,2025ApJ...990..154H,2025arXiv250720232H,2025arXiv251007952F,2025arXiv251020767C}. 
Instead, these source properties are consistent with second (or higher) generation mergers, referred to as hierarchical mergers, with at least one of the black holes (BHs) being the remnant of a previous merger, 
which typically arises from dynamical formation in dense star clusters and active galactic nucleus disks~\citep{2003ApJ...598..419W,2014ApJ...784...71S,2016ARA&A..54..441N,2019ApJ...871...91Z,2020ApJ...903...67M,2021NatAs...5..749G,2019PhRvL.123r1101Y,2023MNRAS.526.6031V,2024A&A...692A..80P,2025PhRvL.134a1401A,2025ApJ...994..261A,2025PhRvD.111j3016L,2025arXiv251203152B,2025ApJ...987...65L,2025ApJ...989L..15Z,2025MNRAS.543.1833M,2025arXiv251022698W}.
In particular, the remnant spin of equal-mass, non-spinning BH mergers is expected to be clustered at $\chi\approx0.7$~\citep{2009PhRvD..79b4003S}. Regarding the mergers of \texttt{2G+1G}, they naturally have an advantage at approximately a $2:1$ mass ratio (higher-generation mergers might have more extreme mass ratios).

Generally, the possibility of a hierarchical merger origin of a GW event, and therefore, its progenitors can be assessed employing hierarchical Bayesian analysis or machine learning techniques by comparing the source parameter distribution with the astrophysical population model.
A recent typical example is GW231123~\citep{2025ApJ...993L..25A} with exceedingly massive components and high spins,
implying that it is highly likely to have originated from hierarchical mergers~(although still debated, see~\citep{2025arXiv250717551L,2025arXiv250810088C,2025arXiv250813412D,2025arXiv250900154P,2025arXiv250908298L,2025arXiv250915619S,2025ApJ...992L..26S,2025arXiv251014363P,2025ApJ...993L..30L,2025PhRvL.135s1401B,2025arXiv251105144H,2025arXiv251113820L,2025ApJ...994L..37K,2025ApJ...994L..54P,2025ApJ...995L...6C,2025arXiv251102691S,2025arXiv251217631G,2025arXiv251217550H}).
However, for a preliminary screening, it is only necessary to determine whether an event has the possibility of having a hierarchical merger origin. The density estimation and comparative analysis of multi-dimensional parameters increase the complexity and computational cost, although they are essential for detailed and accurate analysis~\citep{2021PhRvD.104h4002B,2022ApJ...929L...1B,2023ApJ...945L..18P,2024ApJ...977..220A,2024ApJ...975..117M}.
The merger entropy index, depending primarily on the masses and spins of BBHs to measure the efficiency of entropy transfer in BBH mergers, compresses multidimensional parameters related to mass and spin into one dimension~\citep{2021arXiv211206856H}. It offers an efficient criterion to probe the formation channels of BBH mergers, especially for the hierarchical merger scenario~\citep{2024arXiv241102778C}.
Motivated by this, we timely apply the merger entropy index to investigate the possibility of a hierarchical merger origin of GW241011 and GW241110.

\section{Merger entropy index}

The merger entropy index ($\mathcal{I}_{\rm BBH}$), which is bounded between $[0, 1]$ according to general relativity~\citep{2021arXiv211206856H},
\begin{equation}
\begin{aligned}
    \mathcal{I}_{\rm BBH} =
    \frac{\pi}{9}\,\frac{\Delta S_{\rm BBH}}{S_1+S_2},\,
    \label{eq:I}
\end{aligned}
\end{equation}
where $\Delta S_{\rm BBH}=S_{\rm f}-(S_1+S_2)$ is the increase of entropy during the BBH merger process derived from the second law of thermodynamics; 
and $S_i$ is the entropy of the BH, with the primary, secondary, and remnant BH labeled as $i=1$, $2$, and ${\rm f}$, respectively, in the merger.
The entropy of a Kerr BH~\citep{1973PhRvD...7.2333B} based on the Hawking area theorem~\citep{1971PhRvL..26.1344H} is given by~\citep{1973CMaPh..31..161B}
\begin{equation}
\begin{aligned}
    S = \frac{2\pi G}{c\hbar}\,m^2\,(1+\sqrt{1-\chi^2}),\,
    \label{eq:S}
\end{aligned}
\end{equation}
where $m$ and $\chi$ are the mass and spin of the BH.

Here, we use two types of BBH populations, for comparison, to compute $\mathcal{I}_{\rm BBH}$ in the astrophysical scenario: 
(1) the \textsc{Cluster Monte Carlo} (CMC) catalog~\citep{2020ApJS..247...48K,2022ApJS..258...22R}, which contains a suite of 148 cluster simulations;
and (2) the parametric population (PP) model~\citep{2023PhRvD.107f3007L,2025ApJ...981..177L}, for which we set the synthesis starting from the GWTC-3 distribution~\citep{2023PhRvX..13a1048A} with an escape speed of $\sim$$100\,{\rm km\,s^{-1}}$.
We calculate $\mathcal{I}_{\rm BBH}$ of GW241011 and GW241110 from the GW posterior data sets~\citep{2025ApJ...993L..21A} labeled as ``Mixed.''

We perform the $\mathcal{I}_{\rm BBH}$ comparison between GW sources and population models using two methods, for reference:
(1) the likelihood analysis, which measures the overall degree of matching between the probability density distribution functions of $\mathcal{I}_{\rm BBH}$ for the data and model;
(2) the Kolmogorov-Smirnov (KS) test~\citep{2011MNRAS.415..333P}, which measures the maximum vertical distance between the empirical cumulative distribution functions of $\mathcal{I}_{\rm BBH}$ for the data and model.
We transform the KS statistic ($D$) which reflects the inherent difference to the similarity ($1-D$) between the two distributions following~\citet{2024arXiv241102778C}.
We then normalize the matching-degree and similarity of $\mathcal{I}_{\rm BBH}$ between 1G BBH mergers and hierarchical mergers, under a given GW source and population model.
We note that we neglect observational selection effects, because it has been verified that the inclusion of selection effects does not significantly affect the content of the relevant property range of the GW sources here (see further details in Appendix~D of \citet{2025ApJ...993L..21A}).

\section{Hints to hierarchical mergers}

\begin{figure*}
\centering
\includegraphics[width=15cm]{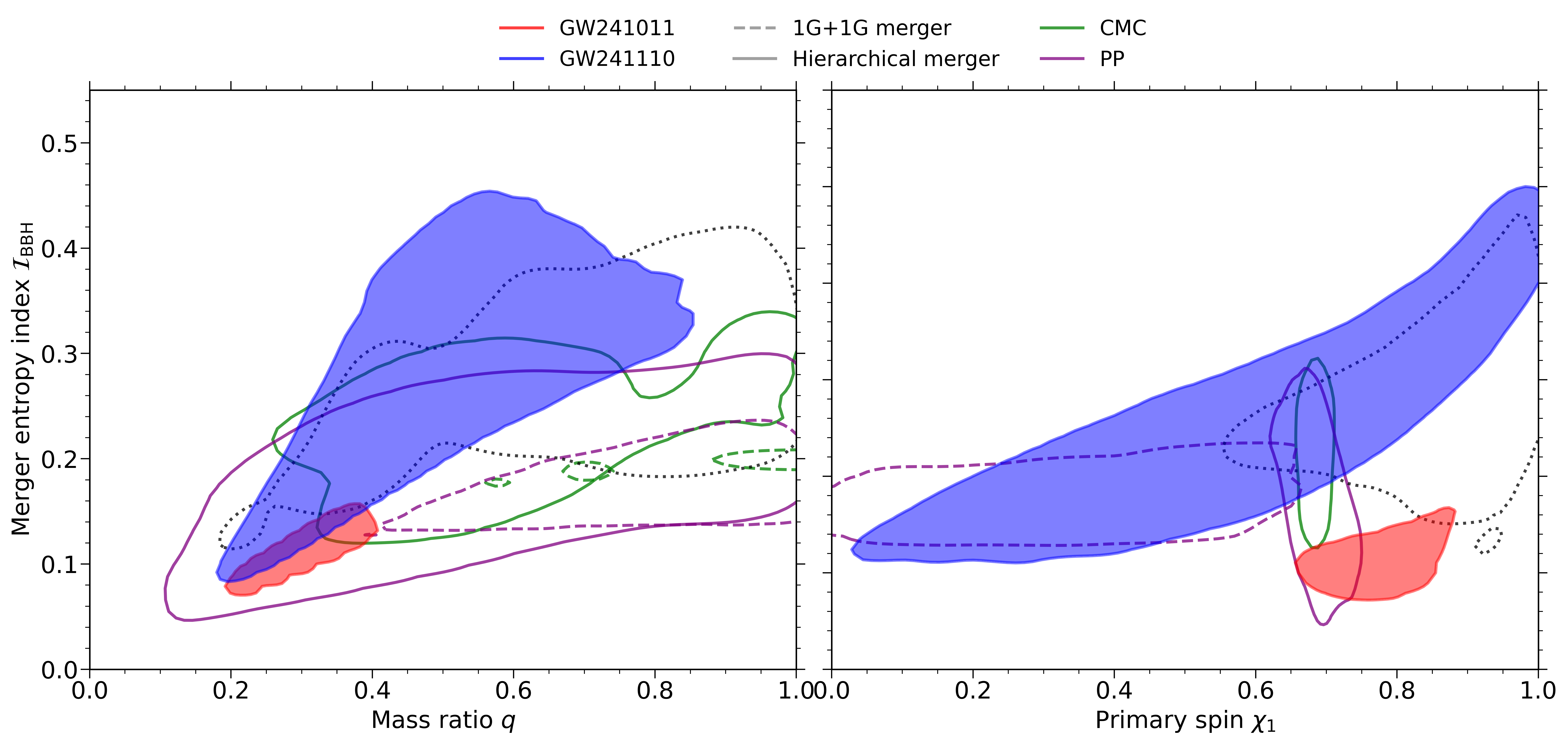}
\caption{
$90\%$ credible bounds on the merger entropy indices, mass ratios (left), and primary spins (right) of GW241011 (red) and GW241110 (blue), compared to expected properties of merging BBHs in dense star clusters from the CMC catalog~\citep{2020ApJS..247...48K,2022ApJS..258...22R} (green) and the PP model~\citep{2023PhRvD.107f3007L,2025ApJ...981..177L} (purple).
Dashed contours correspond to first-generation mergers, while solid contours correspond to hierarchical mergers. Dotted contours represent the properties of GW231123~\citep{2025ApJ...993L..25A} for comparison.
The CMC catalog assumes black holes to be born non-rotating, while the PP model assumes 1G BHs are born with low spins following a beta distribution.
}
\label{fig1} 
\end{figure*}

Figure~\ref{fig1} shows the distribution characteristics of GW241011 and GW241110 in terms of merger entropy index, mass ratio, and primary spin, compared with the 1G BBH mergers and hierarchical mergers from the CMC catalog and the PP model for reference.
The contours in both first-generation mergers and hierarchical mergers of the CMC catalog are consistent with those of the PP model, although the latter does not fully incorporate the complexity of the physical processes governing the formation of higher-generation mergers in dense star clusters (see Discussion section in ~\citet{2025ApJ...981..177L}).
Both $q$-$\mathcal{I}_{\rm BBH}$ and $\chi$-$\mathcal{I}_{\rm BBH}$ contours of GW241011 could exclude its origin as a first-generation merger.
The $q$-$\mathcal{I}_{\rm BBH}$ contours between GW241110 and the first-generation mergers from population models have obvious differences, while their $\chi$-$\mathcal{I}_{\rm BBH}$ contours are chaotic, resulting in no model being able to form a relatively more effective coverage for it.
Regarding GW231123 for comparison, it is similar to GW241011 in exhibiting good differentiation between first-generation mergers and hierarchical mergers like GW241011, although its contour shape is more similar with GW241110.

\begin{figure*}
\centering
\includegraphics[width=15cm]{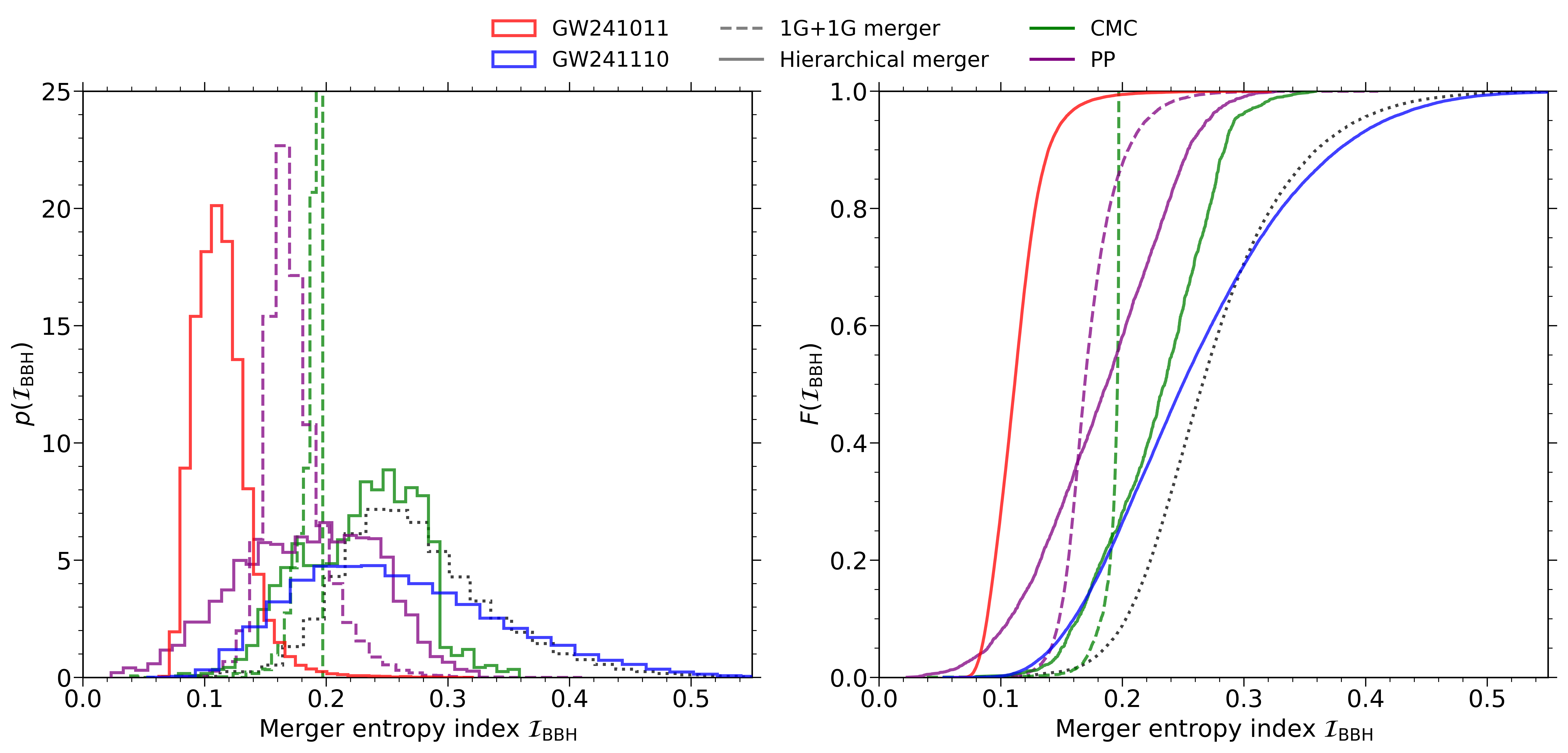}
\caption{
Probability density distributions (left) and cumulative distributions (right) of the merger entropy indices of GW241011 and GW241110, compared to those of merging BBHs in dense star clusters from the CMC catalog~\citep{2020ApJS..247...48K,2022ApJS..258...22R} and the PP model~\citep{2023PhRvD.107f3007L,2025ApJ...981..177L}.
The legend is the same as Figure~\ref{fig1}.
}
\label{fig2} 
\end{figure*}

Figure~\ref{fig2} illustrates the properties of merger entropy indices from GW241011 and GW241110, with respect to the 1G BBH mergers and hierarchical mergers from the CMC catalog and the PP model.
The $\mathcal{I}_{\rm BBH}$ distributions from GW observations and astrophysical models both exhibit the characteristic of a ``unimodal concentratio''.
The $\mathcal{I}_{\rm BBH}$ distribution of first-generation mergers is highly concentrated with the peak of $\sim$$0.2$ for the CMC catalog and $\sim$$0.17$ for the PP model, corresponding to the fastest increase rate of its cumulative distribution.
In contrast, $\mathcal{I}_{\rm BBH}$ of hierarchical mergers is relatively more dispersed and preferentially distributed toward larger values, with the peaks at $\sim$$0.23$ and $\sim$$0.19$, respectively. The $90$\% credible intervals combining the two types of population models are bounded between $[0.14,0.21]$ for first-generation mergers and $[0.09,0.29]$ for hierarchical mergers. 
$\mathcal{I}_{\rm BBH}$ of GW241011 is highly concentrated to significantly lower values than those from first-generation mergers, due to its precisely, extremely unequal-mass components. 
This is because more unequal-mass BBHs are more inclined towards smaller $\mathcal{I}_{\rm BBH}$, while higher spin are inclined towards larger $\mathcal{I}_{\rm BBH}$. 
As a result, $\mathcal{I}_{\rm BBH}$ from GW241110 (and also GW231123~\citep{2025ApJ...993L..25A}) is distributed more evenly with a tendency toward large values, due to, however, their low mass ratios and high spins both having larger uncertainties, which is similar with that of hierarchical mergers.
Consequently, the cumulative trend on hierarchical mergers is similar to that of the observed GW sources (although this is not case for the first-generation mergers from the PP model compared to GW241011), and there is a difference in the upward slope at $\mathcal{I}_{\rm BBH}\sim0.2$.

\begin{table}
\centering
\begin{threeparttable}
\caption{Model comparison results.}
\label{tab}
\setlength{\tabcolsep}{8pt} 
\begin{tabular}{llcccc}
\toprule \toprule
Event & Method & \multicolumn{2}{c}{CMC} & \multicolumn{2}{c}{PP} \\
\cmidrule(lr){3-4} \cmidrule(lr){5-6}
& & 1G & nG & 1G & nG \\
\midrule
\multirow{2}{*}{GW241011}& lik. & $0$ & $1$ & $0$ & $1$ \\
 & KS & $0.30$ & $0.70$ & $0.30$ & $0.70$ \\
 \midrule
\multirow{2}{*}{GW241110} & lik. & $1$ & $0$ & $1$ & $0$ \\
 & KS & $0.26$ & $0.74$ & $0.39$ & $0.61$ \\
\bottomrule
\end{tabular}
\begin{tablenotes}[flushleft]
\footnotesize
\item 
{\bf Note.}
The values are the normalized matching-degrees from the likelihood analysis and similarity from the KS test, of the merger entropy indices from GW sources between first-generation mergers and hierarchical mergers labeled as ``nG.''
\end{tablenotes}
\end{threeparttable}
\end{table}

Table~\ref{tab} presents the model comparison results with the normalized matching-degree (from the likelihood analysis) and similarity (from the KS test; hereafter these two normalized judgment values will be collectively referred to as scores, and the total score from first-generation mergers and hierarchical mergers will be 1) of the merger entropy indices for GW241011 and GW241110, between first-generation mergers and hierarchical mergers.
The results of GW241011 show that both the likelihood and KS tests have higher scores of matching with hierarchical mergers (the likelihood scores of 1 and the KS scores of 0.70), while scores of matching with merging first-generation BBHs are lower (the likelihood scores of 0 and the KS scores of 0.30), suggesting that GW241011 is significantly inclined towards a hierarchical merger origin.

Regarding GW241110, its KS test results also support that this event is more in line with the characteristics of hierarchical mergers (the KS score of 0.74 for the CMC catalog and 0.61 for the PP model). However, the likelihood analysis results of GW241110 with the scores of 1 show that it matches perfectly with first-generation mergers and has a score of 0 with hierarchical mergers. 
This opposite result reflects: (1) the larger distribution uncertainties of GW241110~\citep{2025ApJ...993L..21A}, and (2) the conflict between the ``holistic view'' (the likelihood analysis) and the ``detail view'' (the KS test). 
The likelihood analysis focuses on overall goodness of fit but the distribution peak of first-generation mergers is too sharp, which leads the vast majority of $\mathcal{I}_{\rm BBH}$ values of GW241110 to fall in the high probability distribution region of first-generation mergers, indicating a tendency toward first-generation mergers. 
In contrast, the KS test centers on extreme value differences but a ``heavy tail'' exists in $\mathcal{I}_{\rm BBH}$ of GW241110, which results in the cumulative distribution shape of GW241110 being closest to that of hierarchical mergers, suggesting it matches with hierarchical mergers.
Moreover, the matching trend of the merger mechanism between the CMC and PP models for the two GW events is consistent, highlighting the robustness of the model conclusions~\citep{2020ApJS..247...48K,2022ApJS..258...22R,2023PhRvD.107f3007L,2025ApJ...981..177L}.

\section{Conclusions}\label{sec:conclusions}

We investigate the origin of hierarchical mergers for GW241011 and GW241110~\citep{2025ApJ...993L..21A} using the merger entropy index ($\mathcal{I}_{\rm BBH}$) which measures the efficiency of entropy transfer in BBH mergers~\citep{2021arXiv211206856H,2024arXiv241102778C}.
We adopt two types of BBH populations, the CMC catalog~\citep{2020ApJS..247...48K,2022ApJS..258...22R} and the PP model~\citep{2023PhRvD.107f3007L,2025ApJ...981..177L} for comparison, and the same results from the two models highlight the consistency of the model conclusions.
We perform the $\mathcal{I}_{\rm BBH}$ comparison between GW sources and population models using two methods, the likelihood analysis and the KS test, respectively.

The analysis results of GW241011 show that both the likelihood and KS tests have a higher degree of matching with hierarchical mergers (see Table~\ref{tab}), suggesting that GW241011 is strongly favored to originate from the hierarchical merger mechanism.
The KS test results of GW241110 support that it is consistent with the characteristics of hierarchical mergers, while its likelihood analysis results indicate that GW241110 might match those o first-generation mergers.
This is because of the large uncertainty of GW241110 due to its lower SNR and the different emphases of the statistical methods, highlighting that these effects should be taken into account when interpreting this event in terms of a hierarchical merger origin.

\section{Acknowledgments}
This work is supported by National Key R$\&$D Program of China (2020YFC2201400). 
This research has made use of data or software obtained from the Gravitational Wave Open Science Center (\url{https://gwosc.org}), a service of the LIGO Scientific Collaboration, the Virgo Collaboration, and KAGRA. This material is based upon work supported by NSF's LIGO Laboratory which is a major facility fully funded by the National Science Foundation, as well as the Science and Technology Facilities Council (STFC) of the United Kingdom, the Max-Planck-Society (MPS), and the State of Niedersachsen/Germany for support of the construction of Advanced LIGO and construction and operation of the GEO600 detector. Additional support for Advanced LIGO was provided by the Australian Research Council. Virgo is funded, through the European Gravitational Observatory (EGO), by the French Centre National de Recherche Scientifique (CNRS), the Italian Istituto Nazionale di Fisica Nucleare (INFN) and the Dutch Nikhef, with contributions by institutions from Belgium, Germany, Greece, Hungary, Ireland, Japan, Monaco, Poland, Portugal, Spain. KAGRA is supported by Ministry of Education, Culture, Sports, Science and Technology (MEXT), Japan Society for the Promotion of Science (JSPS) in Japan; National Research Foundation (NRF) and Ministry of Science and ICT (MSIT) in Korea; Academia Sinica (AS) and National Science and Technology Council (NSTC) in Taiwan.
This analysis was made possible following software packages:
NumPy~\citep{harris2020array}, 
SciPy~\citep{2020SciPy-NMeth}, 
Matplotlib~\citep{2007CSE.....9...90H}, 
IPython~\citep{2007CSE.....9c..21P},
and seaborn~\citep{Waskom2021}.

\newcommand{\jcap}{J. Cosmol. Astropart. Phys.}
\newcommand{\physrep}{Phys. Rep.}
\newcommand{\mnras}{Mon. Not. R. Astron. Soc.}
\newcommand{\araa}{Annu. Rev. Astron. Astrophys.}
\newcommand{\aap}{Astron. Astrophys}
\newcommand{\aj}{Astron. J.}
\newcommand{\plb}{Phys. Lett. B}
\newcommand{\apjs}{Astrophys. J. Suppl.}
\newcommand{\app} {Astropart. Phys.}
\newcommand{\apjl}{Astrophys. J. Lett}
\newcommand{\pasp}{Publ. Astron. Soc. Pac.}

%

\end{sloppypar}\end{document}